\documentclass[twocolumn,aps,pra,notitlepage,10pt,superscriptaddress]{revtex4-1}
\usepackage[utf8]{inputenc}
\usepackage{graphicx,braket,amsmath,amssymb,amsfonts,graphics,color,nicefrac}
\usepackage{verbatim,pst-all,pst-plot,flushend,tabularx,multirow,booktabs,threeparttable}
\usepackage[T1]{fontenc}
\usepackage{libertine,epstopdf,array,pbox,makecell,subfigure,enumitem,subfiles,listings,fancybox,stmaryrd}

\newcolumntype{P}[1]{>{\centering\arraybackslash}p{#1}}

\usepackage{titlesec}
\titlespacing\section{0pt}{12pt plus 4pt minus 2pt}{0pt plus 2pt minus 2pt}
\titlespacing\subsection{0pt}{10pt plus 4pt minus 2pt}{0pt plus 2pt minus 2pt}
\titlespacing\subsubsection{0pt}{10pt plus 4pt minus 2pt}{0pt plus 2pt minus 2pt}

\begin{document}
\title{Decoding surface code with a distributed neural network based decoder}
\author{Savvas Varsamopoulos}
\author{Koen Bertels}
\author{Carmen G. Almudever}
\affiliation{Quantum Computer Architecture Lab, Delft University of Technology, The Netherlands}
\affiliation{QuTech, Delft University of Technology, P.O. Box 5046, 2600 GA Delft, The Netherlands \\ Email: S.Varsamopoulos@tudelft.nl}

\begin{abstract}
There has been a rise in decoding quantum error correction codes with neural network based decoders, due to the good decoding performance achieved and adaptability to any noise model. However, the main challenge is scalability to larger code distances due to an exponential increase of the error syndrome space. Note that, successfully decoding the surface code under realistic noise assumptions will limit the size of the code to less than 100 qubits with current neural network based decoders.

Such a problem can be tackled by a distributed way of decoding, similar to the Renormalization Group (RG) decoders. In this paper, we introduce a decoding algorithm that combines the concept of RG decoding and neural network based decoders. We tested the decoding performance under depolarizing noise with noiseless error syndrome measurements for the rotated surface code and compared against the Blossom algorithm and a neural network based decoder. We show that similar level of decoding performance can be achieved between all tested decoders while providing a solution to the scalability issues of neural network based decoders.
\end{abstract}
\maketitle

\section{Introduction}
\label{sec:Introduction}

\textbf{\textit{Quantum error correction (QEC)}} is for now considered to be the most time and resource consuming procedure in quantum computation. However, the way that quantum computing is currently envisioned, QEC is necessary for reliable quantum computation and storage. The need for QEC arises from the unavoidable coupling of the quantum system with the environment, which causes the qubit state to be altered (decohere). Altering the quantum state is perceived as errors generated in the quantum system. Through active error correction and fault-tolerant mechanisms, that control error propagation and keep the error rates low, we can have the error-free desired state. Note that, in \textbf{\textit{fault-tolerant}} techniques, errors can occur in the quantum system, but do not affect the quantum state in a catastrophic manner \cite{nielsen}.

A critical sub-routine of QEC is \textbf{\textit{decoding}}. Decoding involves the process of identifying the errors that occur in the quantum system and proposing corrections that keep the quantum state error-free. The importance of high speed and accurate decoding lies in the fact that the time budget allowed for error correction is small, since qubits lose their state rapidly. Therefore, if the process of decoding exceeds the error correction time budget, errors will accumulate to the point that the error-free state cannot be retrieved.

Various classical decoding algorithms have been proposed over the years with a few examples of classical decoding algorithms being the Blossom algorithm \cite{edmonds, kolmogorov, fowler_o1, fowler_opt_comp}, the maximum-likelihood algorithm \cite{Bravyi_Suchara_Vargo} and the Renormalization Group (RG) algorithm \cite{duclos, duclos1}. Recently, there is an increase in the development of \textbf{\textit{neural network based decoders}} that either consist exclusively of neural networks \cite{Torlai, Krastanov} or a classical module working together with neural networks \cite{Savvas, Baireuther, Chamberland2018, Ni,Davaasuren_2018}. Neural network based decoders exist with different designs in the way the decoding is performed and a variety of types of neural networks has been explored, like Feed-forward, Recurrent and Convolutional neural networks. 

\begin{figure}[hbt]
\centering
\scalebox{0.65}{
\includegraphics[width=\columnwidth]{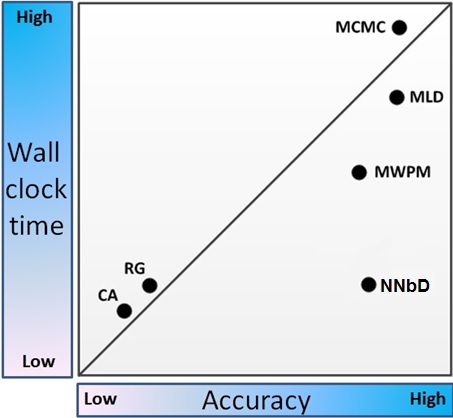}}
\caption{Abstract comparison between decoding performance and execution time of various decoding algorithms}
\label{fig:perf_vs_time}
\end{figure}

In Figure \ref{fig:perf_vs_time} we present an abstract comparison between various decoding algorithms based on their decoding performance (Accuracy) and their execution time (Wall clock time), namely the Markov Chain Monte Carlo (MCMC) \cite{MCMC}, the Maximum Likelihood Decoder (MLD) \cite{Bravyi_Suchara_Vargo}, the Minimum Weight Perfect Matching (MWPM) \cite{edmonds, kolmogorov} that Blossom algorithm is based on, the Neural Network based Decoder (NNbD) \cite{savvas_nn_designs}, the Renormalization Group (RG) \cite{duclos1} and the Cellular Automaton (CA) \cite{CA}. Decoding performance is typically calculated as the ratio of the number of logical errors created out of the decoder corrections over the number of error correction cycles run to accumulate these errors. Execution time is defined as the time spent from the moment that the input data arrive at the decoder until the time that the decoder proposes the corrections. As can be seen from Figure \ref{fig:perf_vs_time}, neural network based decoders can reach equivalent decoding performance as classical algorithms while requiring smaller execution time. This is the main reason that neural network based decoders are explored and various designs have been proposed recently. However, the main issue with such decoders is that scaling to larger quantum systems will be significantly harder compared to classical decoders, due to the required training process of the neural network. As the size of the system increases, more training samples need to be collected and then the neural network has to be trained based on them. The main challenge of NNbDs is that in order to reach similar decoding performance to classical algorithms as the quantum system is increasing, the amount of samples required to be collected increases in an exponential way, which makes the training harder and slower. 

In this work, we will present a neural network based decoder that performs decoding in a distributed fashion, therefore providing a solution for the issue of decoding large codes. We should mention that there exist classical algorithms that perform decoding in a distributed way, as can be found in \cite{duclos1} and \cite{fowler_o1}, but in this paper we will provide a different approach of the distributed decoding concept. In \cite{duclos1}, the original idea of RG decoding approach is described and tested. RG decoding is based on the division of the code into small tiles, in which a given number of physical qubits are included and error probabilities about the physical qubits inside all tiles are calculated. Then, these tiles are grouped into larger tiles and the error probabilities about the qubits are updated. This procedure is continued until only a single tile has remained containing all the physical qubits of the system. Based on the updated error probabilities of the largest tile, the decoder can propose a set of corrections. In \cite{fowler_o1}, a distributed decoding approach is described, where the code is divided into small tiles. However, in this case Blossom algorithm is used to decode each tile and based on the result of it and the neighboring information between the tiles, the decoder can propose corrections for the whole code. Each tile is monitored by an Application-Specific Integrated Circuit (ASIC), which is dedicated for the tile.

In our strategy, the code is divided into small overlapping regions, referred to as \textbf{\textit{overlapping tiles}}, where local information about errors on physical qubits is obtained. Then, this local information is combined and a decoding for the whole code is obtained. We compare our algorithm to the unoptimized version of Blossom algorithm \cite{edmonds, kolmogorov} and argue about the decoding performance achieved. Furthermore, we will provide reasoning for the potential high level of parallelization of our algorithm that will be suitable for a high speed hardware implementation without loss of decoding performance. Also, the problem of the exponential increase of the error syndrome space is mitigated, since it is controlled by the selection of the size of the decoded regions. This allows neural network based decoders to successfully decode larger codes.

The rest of the paper is organized in the following way: in sections ~\ref{sec:QEC}, and ~\ref{sec:RG} we give a short introduction in quantum error correction and the concept of RG decoding, respectively. In section ~\ref{sec:Over_tiles}, we present the design of the distributed neural network based decoder and in section ~\ref{sec:Results}, we provide the results in terms of decoding performance. Finally, in section ~\ref{sec:Conclusions}, we draw our conclusions about the distributed decoding approach.

\section{Quantum error correction}
\label{sec:QEC}

Quantum computation is error prone due to the fragility of the qubits, which lose their coherence through their interaction with the environment. Furthermore, quantum operations are still imperfect, altering the quantum state in unpredictable ways. These alterations are interpreted as errors in the quantum system, which are discretized into Pauli errors in order to be corrected in an easier way.

Quantum error correction involves an encoding process of the quantum information into multiple qubits and a decoding process that identifies and counteracts the noise that is inserted in the quantum system. Many unreliable \textbf{\textit{physical qubits}} are encoded, similarly to classical error correction, to one more reliable qubit, known as \textbf{\textit{logical qubit}}. There are many ways that encoding can be achieved, these encoding schemes are also known as quantum error correcting codes \cite{Landahl_color, fowler_sc, Suchara_sub, Bravyi_sub, Bombin_sub, Bravyi_3_check}, but we are focusing on the \textbf{\textit{surface code}} \cite{gottesman, Kitaev_orig}. 

Logical qubits are used both for quantum computation and memory, however, errors occur at the physical level. Therefore, a decoding process that will identify the errors on the physical qubits is required. At the end of the decoding process, corrections against identified errors are proposed by the decoder.

\subsection{Surface code}
\begin{figure}[htb]
\centering
\includegraphics[width=\columnwidth]{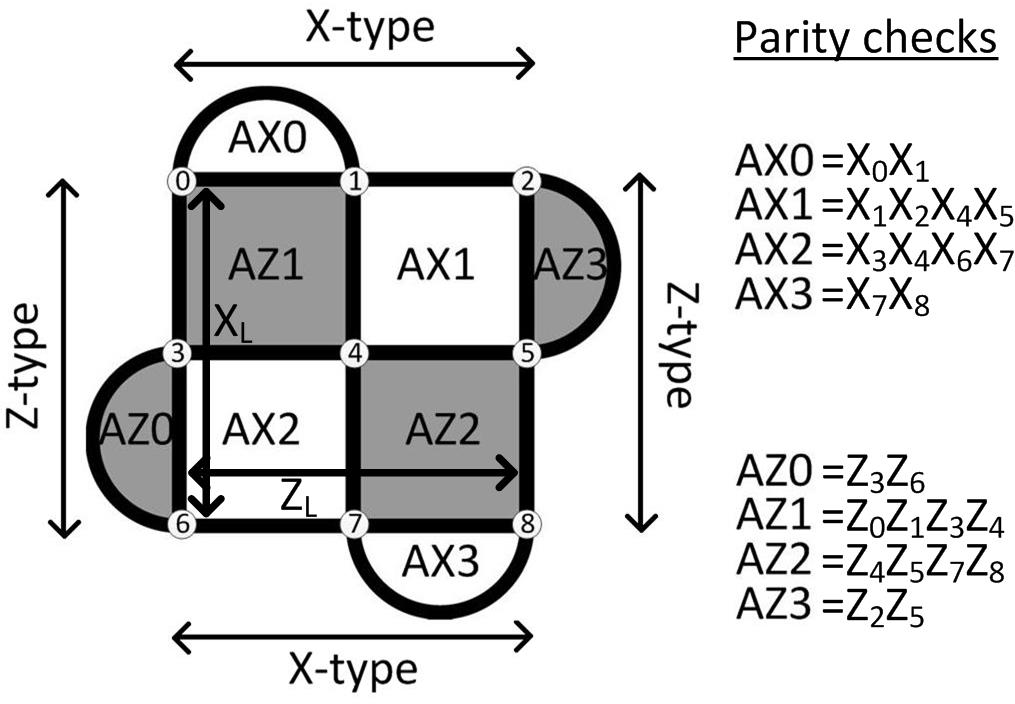}
\caption{Rotated surface code describing 1 logical qubit that consists of 17 physical qubits. The 9 qubits that are at the corners of the tiles (0-8) are data qubits and the 8 qubits that are inside the tiles (AXi, AZi) are ancilla qubits. The parity-checks of the code are shown on the right side.}
\label{fig:d_3_stabs}
\end{figure}

The surface code is a topological stabilizer code with simple structure, local interactions and high level of protection against errors \cite{dklp,Raussendorf, fowler_high_thres,Wang_high,fowler_clas_proc,Bombin_2009,Bombin_2011,Bravyi98,fowler_sc}. A logical qubit in the surface code includes two types of physical qubits, namely the \textbf{\textit{data qubits}}, which store quantum information, and ancillary or \textbf{\textit{ancilla qubits}}, which can be used to find errors on the data qubits. The smallest version of a planar surface code \cite{Bravyi98,freedman2001projective} which requires the least amount of physical qubits, known as the \textbf{\textit{rotated surface code}} \cite{Horsman2012}, is presented in Figure \ref{fig:d_3_stabs}.

A logical qubit is defined by its \textbf{\textit{logical operators}} ($\bar{X_L}$, $\bar{Z_L}$), which are responsible for logical state changes. Any operator of the form $X^{\otimes n}$ or $Z^{\otimes n}$ that forms a chain which spans two boundaries of the same type, can be considered as a logical operator, with $n$ being the amount of data qubits included in the logical operator. The operator with the smallest $n$ is always selected, however as can be seen from Figure \ref{fig:d_3_stabs} there are multiple logical operators with $n=3$, which is the smallest $n$ for this code. Any one of them can be selected without further assumptions. For example, a valid $\bar{X_L}$ could be $X_{0}X_{3}X_{6}$ and a valid $\bar{Z_L}$ could be $Z_{6}Z_{7}Z_{8}$.

The level of protection against errors is usually described with the metric known as \textbf{\textit{code distance}}. Code distance, $d$, is calculated as the minimum number of physical operations required to change the state of the logical qubit \cite{Terhal, beginners}. Therefore, for the logical qubit of Figure \ref{fig:d_3_stabs} the code distance would be 3.

The relation between the code distance and the errors that can be successfully corrected is given by:

\begin{equation}
	\text{weight of error}=\lfloor \frac{d-1}{2} \rfloor
\end{equation}

According to eq. 1, for a $d=3$ surface code, all single errors (weight=1) are going to be successfully corrected.

\begin{figure}[htb]
\centering
\includegraphics[width=\columnwidth]{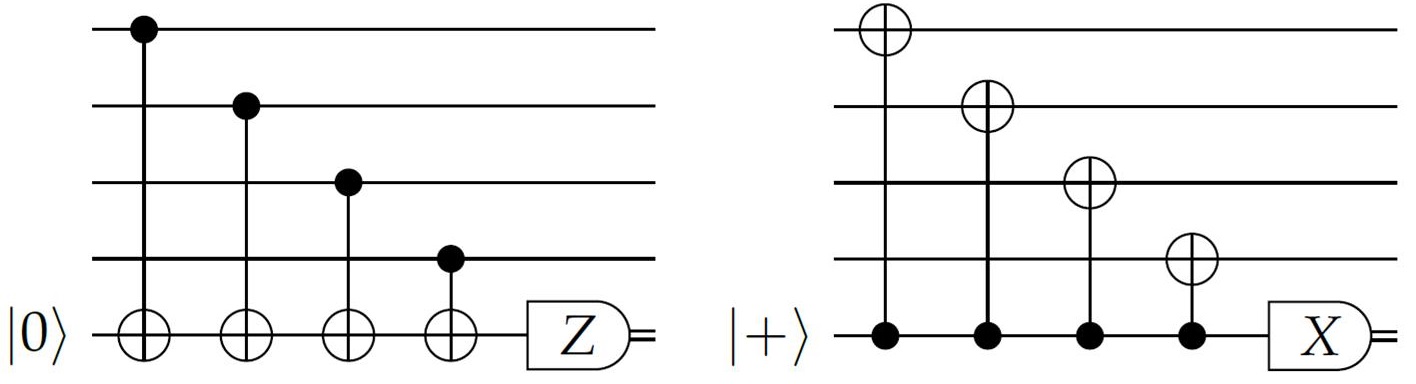}
\caption{Left: Circuit for Z-type ancilla. Right: Circuit for X-type ancilla.}
\label{fig:SCcycle}
\end{figure}

Since the errors are discretized into bit- and phase-flip errors, it is sufficient to only have two types of ancilla qubits, a Z-type for detecting bit-flips and a X-type for detecting phase-flips. Each ancilla qubit that resides inside a tile and interacts with 4/2 neighboring data qubits to perform a parity-check operation. We provide the parity-checks for a d=3 rotated surface code in Figure \ref{fig:d_3_stabs}, as obtained by running the circuits depicted in Figure \ref{fig:SCcycle}. These circuits are run in parallel and constitute a surface code (error correction) cycle. Both circuits consist of: initialization of the ancilla qubit, followed by a series of \textsc{CNOT} gates between the ancilla and the data qubits, followed by ancilla measurement.

The result of the ancilla measurement is a binary value that indicates whether the value of the parity-check measured, is the same as the one of the previous error correction cycle or not. When a parity-check returns a different value between two consecutive surface code cycles, it is referred to as a \textbf{\textit{detection event}}. By running the circuits of Figure \ref{fig:SCcycle}, we obtain the values for all parity-checks and infer what errors have occurred. Gathering all parity-check values out of a single surface code cycle forms the \textbf{\textit{error syndrome}}.

\subsection{Error decoding}
\begin{figure}[htb]
\centering
\scalebox{0.5}{
\includegraphics[width=\columnwidth]{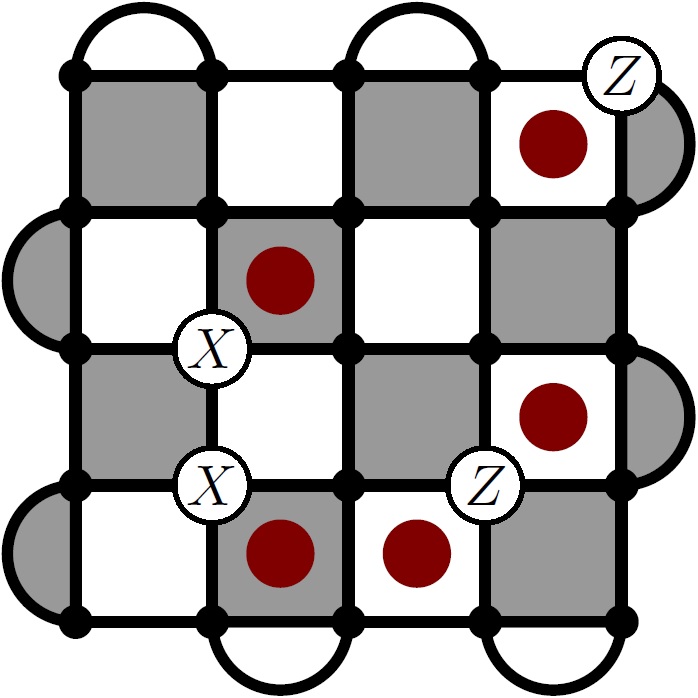}}
\caption{Rotated surface code with code distance 5. Errors are denoted on top of the data qubits with X or Z and detection events corresponding to these errors are shown with red dots.}
\label{fig:errors_2D}
\end{figure}

A single data qubit error will cause two neighboring parity-checks to indicate two detection events (Z error in the bottom of the lattice in Figure \ref{fig:errors_2D}), unless the error occurs at the corner of the lattice which will lead to only one parity-check indicating one detection event (Z error in the top corner of the lattice in Figure \ref{fig:errors_2D}). Multiple data qubit errors that occur near each other form chains of errors (X errors in Figure \ref{fig:errors_2D}), which causes only two detection events located at the parity-checks existing at the endpoints of the error chain \cite{dklp,Terhal,fowler_sc}.

\begin{figure}[htb]
\centering
\scalebox{0.6}{
\includegraphics[width=\columnwidth]{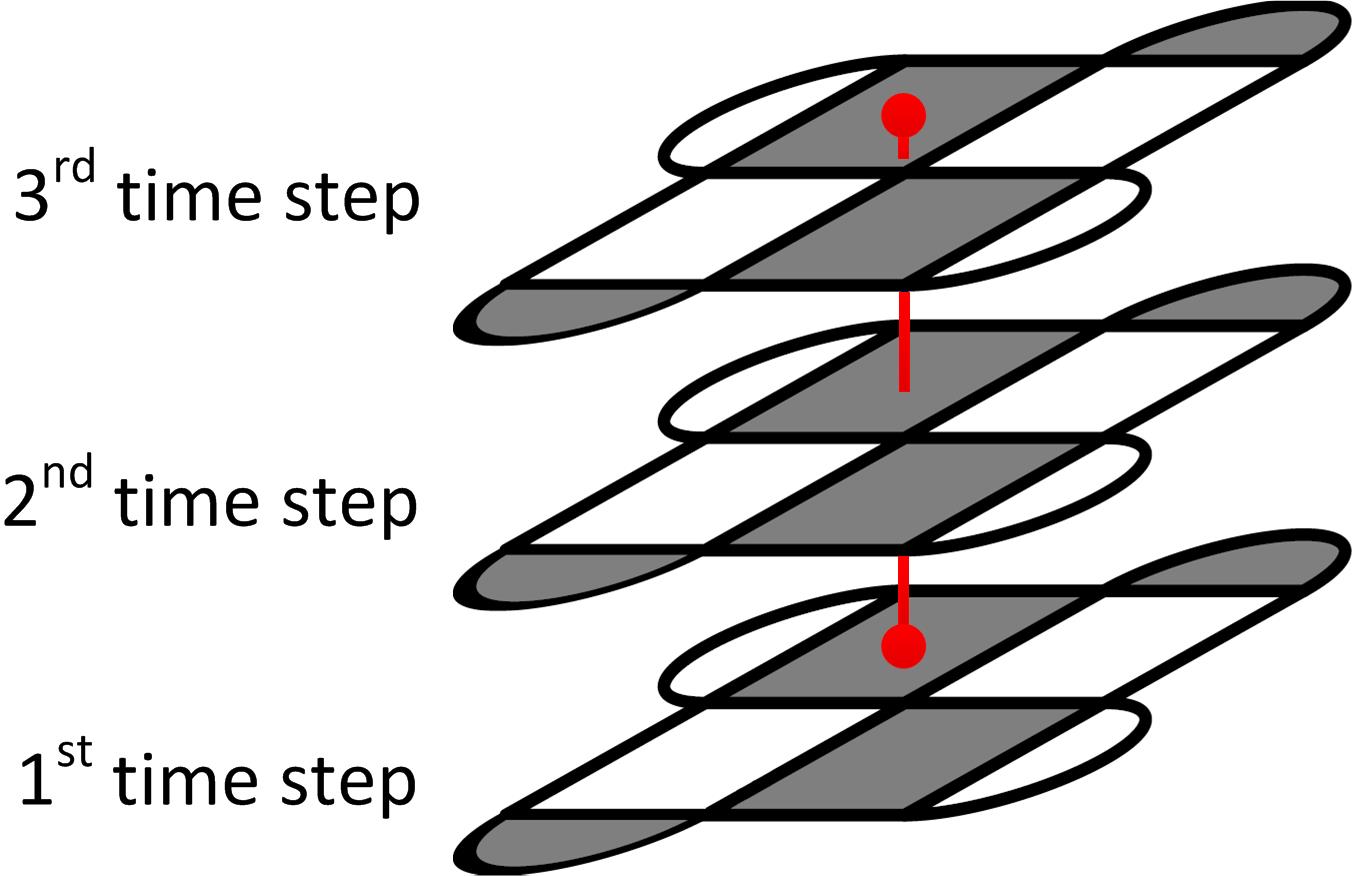}}
\caption{Rotated surface code with code distance 3 at consecutive time steps. Alternating pattern on the measurement value of the same parity-check, indicates the presence of a measurement error.}
\label{fig:error_3D}
\end{figure}

In addition, the measurement process is also imperfect, which leads to different type of errors. When a measurement outcome is misinterpreted, a correction might be applied where no error existed and vice-versa. The way that a measurement error is observed is by comparing the measurement values of multiple consecutive surface code cycles for the same parity-check, as presented in Figure \ref{fig:error_3D}. 

In the case where the error probability for a data qubit error is equal to the error probability for a measurement error, $d$ surface code cycles are deemed enough to successfully identify measurement errors \cite{Raussendorf_2007}. When a measurement error is successfully identified, no correction is required.

Thus, through observation of the parity-checks throughout multiple surface code cycles, identification of errors is made in space (data errors) and in time (measurement errors). The decoder, which is the module responsible for analyzing the detection events and producing corrections against the errors that have occurred, receives the error syndrome out of one or multiple surface code cycles and produces a set of corrections to be applied. 

However, totally suppressing the noise is unfeasible, since the decoder might misinterpret the information coming from the error syndrome. The main reason for such misinterpretations, comes from the fact that the surface code is a degenerate code. This degeneracy means that different sets of errors create the same error syndrome. Therefore, based on the physical error rate of the quantum operations, different sets of errors are more likely than others. This puts an extra assumption to the decoder, since it should output different corrections based on the error probability. Based on all these reasons, it is evident that no decoder can perfectly suppress all noise.

\subsection{Decoding algorithms}
The main parameters that define a good decoder are the decoding performance, the ability to efficiently scale to large code distances and the execution time. There exist decoders that can reach good decoding performance, enough to make fault-tolerant quantum computing possible. Some of the classical algorithms are the \textbf{\textit{maximum-likelihood algorithm}} \cite{Bravyi_Suchara_Vargo}, the \textbf{\textit{Blossom algorithm}} \cite{edmonds, kolmogorov, fowler_o1}, and the \textbf{\textit{Renormalization Group (RG)}} algorithm \cite{duclos, duclos1}. The maximum-likelihood algorithm investigates the most probable error that has occurred that produces the observed error syndrome. This process can reach high decoding accuracy but is extremely time consuming especially as the code distance increases. The execution time scales as $O(n\chi^3)$, with $\chi$ being an approximation parameter, as given in \cite{Bravyi_Suchara_Vargo}. The Blossom algorithm can reach slightly lower decoding performance than the maximum-likelihood decoder, but still good enough to be used in experiments. The execution time scales linearly with the number of qubits \cite{fowler_opt_comp}, but still might not meet the small execution time requirements of contemporary experiments. However, there exist an optimized version of the Blossom algorithm that claims a constant average processing time per detection round, which requires dedicated hardware \cite{fowler_o1}. Renormalization Group decoding provides a good solution for the decoding of large quantum systems, because decoding is performed in a local manner through distributed regions throughout the lattice. The RG algorithm can be highly parallelized and the scaling is reported to be $log(l)$, for an $l$x$l$ code \cite{duclos1}. However, the decoding accuracy is not as good as the other two algorithms. Neural network based decoders with a large variety of designs \cite{Torlai, Krastanov,Savvas, Baireuther, Chamberland2018,Ni,savvas_nn_designs,Maskara_2018,Darmawan,Sweke} have been recently suggested that report similar or better decoding performance than Blossom and RG decoders, making them a potential candidate for decoding.

Currently, the time budget for error correction and decoding is small for most qubit technologies, due to the erroneous nature of the qubits and the imperfect application of quantum operations. Therefore, a high speed version of a decoder would be necessary. This requirement lead us to neural network based decoders which are shown to have constant execution time after being trained. However, in order to run complex algorithms many qubits are required and as mentioned earlier scaling to large code distances with neural network based decoders is extremely hard, since the amount of data required to train the algorithm grow exponentially with the number of qubits. 

In this paper, we will present a neural network based decoder that exploits the concept of distributed decoding, in a similar way to RG decoding and the parallel approach of \cite{fowler_o1}. Based on such a distributed way of decoding, we limit the amount of training data required, making the distance of the code irrelevant.

\section{RG decoding}
\label{sec:RG}

Our previous efforts were mainly focused on developing neural network based decoders that can achieve better decoding performance than classical decoding algorithms and report a constant execution time for each code distance for all range of physical error probabilities, which scales linearly with the code distance \cite{savvas_nn_designs}. However, good decoding performance was harder to achieve as the code distance increased. The main problem was the exponential increase of the error syndrome space, which required an immensely large number of training samples in order for the decoder to achieve similar performance to the classical decoding algorithms for d$>$9. We provide the size of the training datasets used for the code distances investigated in \cite{savvas_nn_designs} for the depolarizing error model in Table \ref{table:data_size}.

\begin{table}[ht]
\caption{Size of training datasets}\label{table:data_size}
\centering
\begin{tabular}{ | c | c | c |} \hline
\text{code distance} & \text{selected dataset size} & \text{full dataset size} \\ \hline
\text{d=3} & 256 & $2^8$\\ \hline
\text{d=5} & $6*10^5$ & $2^{24}$\\ \hline
\text{d=7} & $5*10^6$ & $2^{48}$\\ \hline
\text{d=9} & $2*10^7$ & $2^{80}$\\ \hline
\end{tabular}
\end{table}

A way that the error space can be limited, is through a distributed way of decoding similar to the RG algorithm. By dividing the code in small regions which are going to provide individual information about decoding every region of the code, the decoder can have enough information about decoding the whole code. Limiting the region that we want to locally decode, the error syndrome space is also limited, allowing us to increase the distance of the code without changing the decoding of each region.

\begin{figure}[ht]
\centering
\scalebox{0.8}{
\includegraphics[width=\columnwidth]{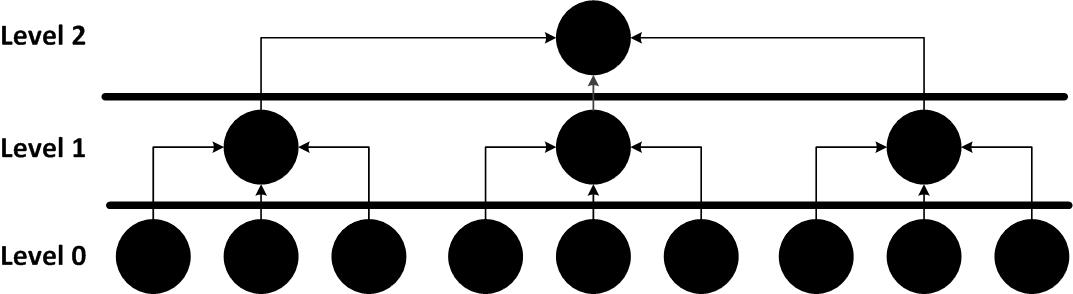}}
\caption{Encoding levels of a concatenated code. At level 0 there are nine qubits, that are encoded in three qubits at level 1 and these qubits are encoded in one qubit at level 2. Arrows show the information flow.}
\label{fig:concatenation}
\end{figure}

RG decoding is similar to decoding concatenated codes, which have various levels of encoding, as can be seen at Figure \ref{fig:concatenation}. 

In these codes, decoding is achieved by passing the error information concerning the qubits from the lower level to the higher level. The information about errors is updated throughout the encoding levels. The decoding occurs at the last encoding level and a final decision about the logical state is made.

\begin{figure}[htb]
\centering
\scalebox{0.6}{
\includegraphics[width=\columnwidth]{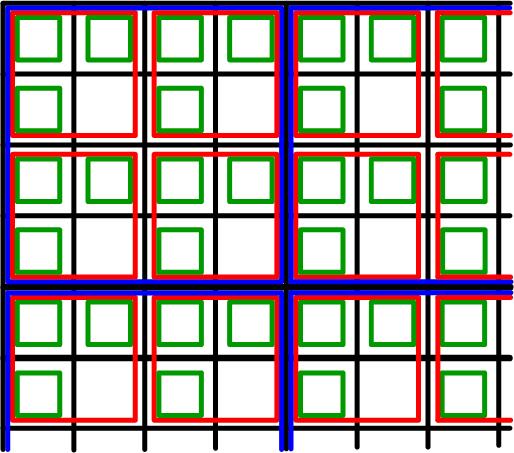}}
\caption{Tile segmentation that represents the levels of concatenation in a concatenated code. The smallest level of concatenation is represented by the green tiles, the next level of concatenation is represented by the red tiles, the following level of concatenation is represented by the blue tiles, etc.}
\label{fig:RG_blocks}
\end{figure}

The strategy of RG decoding can be described according to Figure \ref{fig:RG_blocks}. At first, the lattice is cut in small (green) tiles and the probability of an error occurring in all qubits included in that tile is evaluated. After gathering the updated error probabilities in the green tiles, the lattice is cut into bigger (red) tiles and the error probability of all qubits included in that tile is evaluated. This process is continued until there is only one tile left that includes all qubits in the code.

The same approach can be applied to surface code. However, the challenge here is that the parity-checks cannot be broken down into constant size tiles in a way that every parity-check corresponds to a single tile. Therefore, we need to use overlapping tiles, which will always include whole parity-checks of the code in a single tile. The boundary qubits that belong to neighboring tiles are treated as independent variables on each tile and the error probability for the same qubit is different depending on the tile. The way that the error probabilities are usually calculated is by belief propagation \cite{duclos, duclos1} in the RG approach.

We decided to use the idea of overlapping tiles, but follow a different approach than the RG algorithm as we will explain in the following section.

\section{Distributed decoding with overlapping tiles}
\label{sec:Over_tiles}

We developed a neural network based decoder that performs distributed decoding based on the concept of RG decoders. As mentioned, the main idea behind this algorithm is to make neural network based decoders able to successfully decode large code distances. By restricting the decoding in small regions (tiles) of the lattice, the decoder does not have to explore a large error syndrome space, rather just decode every small tile and then combine the information out of all tiles.

The main difference between a distributed neural network based decoder and the RG decoder is that the former only has one level of concatenation. Instead of moving from smaller tile to bigger tile until the whole lattice is a single tile, we segment the lattice into small equally sized tiles that are overlapping with each other, so that each tile includes whole parity-checks of the code. Then, we obtain error information from each individual tile and combine the information out of all tiles to get the error information for the whole lattice. In this case, there is no need to calculate the error probability of all qubits and forward it to the next level of concatenation, rather find a way to combine the information arising from the each tile.

\begin{figure}[htb]
\centering
\scalebox{0.65}{
\includegraphics[width=\columnwidth]{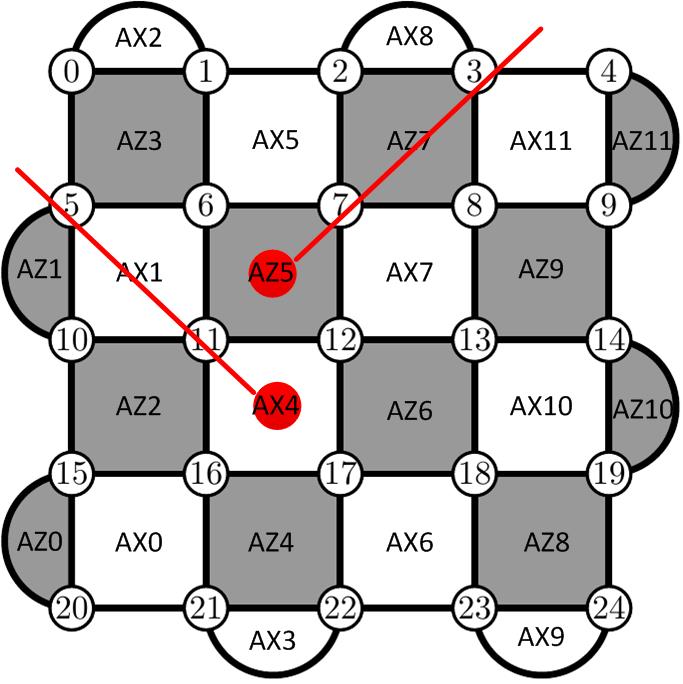}}
\caption{Description of the simple decoder operation for the rotated surface code with distance 5. Detection events are presented with the red dots. Red lines indicate which data qubits are going to be corrected.}
\label{fig:simple_dec}
\end{figure}

In order to decode based on the distributed decoding approach, we will use the same two-module decoder as was presented in \cite{savvas_nn_designs}. Our decoding algorithm consists of two modules, a classical decoding module that we call \textbf{\textit{simple decoder}} and a neural network. The simple decoder provides a naive decoding for the whole lattice, in which a chain is created between each detection event and its closest boundary of the same type. The corrections arising from the simple decoder occur in the data qubits underneath the chain. An example is provided in Figure \ref{fig:simple_dec}, where AZ5 and ancilla AX4 have indicated the presence of an error in their proximity. The proposed corrections of the simple decoder will be Z5, Z11 arising from ancilla AX4 and X3, X7 arising from ancilla AZ5.

The simple decoder receives the error syndrome for the whole lattice and provides a set of corrections for the whole lattice. This is a fast process since the corrections arising from each detection event are independent from the corrections arising from other detection events, therefore can be parallelized. However, the simple decoder cannot yield high decoding accuracy on its own, due to its simplistic design.

That is why we also include the neural network that will work as a supervisor to the simple decoder. More accurately, the neural network will be trained to identify for which error syndromes the simple decoder will lead to a logical error. In the case where a logical error will be created out of the simple decoder corrections, the neural network will output the appropriate logical operator that will cancel the logical error out. As we showed in \cite{savvas_nn_designs}, the combination of these two modules will provide high decoding performance.

In order to train the neural network, we create a training dataset by running surface code cycles and storing the error syndrome and the corresponding logical state of the logical qubit after the corrections of the simple decoder are applied. The size of the training dataset varies based on the code distance and the error model. For more information about all the parameters that affect the dataset, we refer the reader to our previous work \cite{savvas_nn_designs}. 

\begin{figure}[htb]
\centering
\scalebox{0.95}{
\includegraphics[width=\columnwidth]{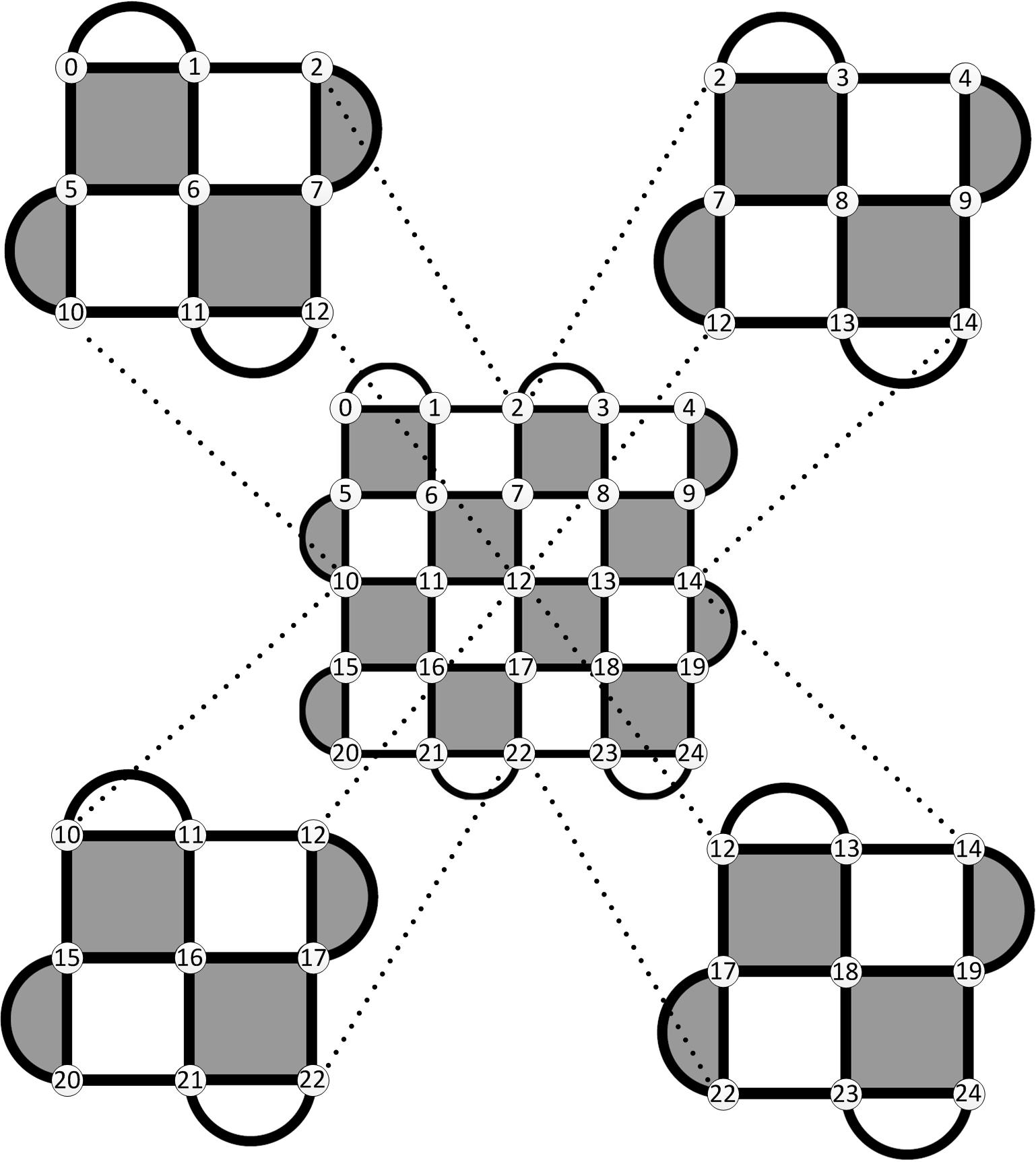}}
\caption{Segmentation of a d=5 rotated surface code into four overlapping tiles of d=3 rotated surface codes.}
\label{fig:4_d=3_in_d=5}
\end{figure}

In Figure \ref{fig:4_d=3_in_d=5}, we provide an example of the segmentation of a d=5 rotated surface code into four overlapping tiles of d=3 rotated surface codes.

As can be seen from Figure \ref{fig:4_d=3_in_d=5}, each parity-check is included in at most two tiles. The error syndrome obtained for the whole lattice (d=5) is broken down into parts of the error syndrome that correspond to each small tile (d=3). The error syndrome out of one surface code cycle consists of 24 bits, due to the 24 parity-checks of the d=5 code. The error syndrome will be cut into smaller parts of the initial error syndrome that fit the d=3 tiles. Due to inclusion of the shared parity-checks, the bits that are available out of the four d=3 tiles are now 32. Each error syndrome of the d=3 tile corresponds to a part of the complete error syndrome. The error probabilities of the logical state, $\text{Prob(}I\text{)}$, $\text{Prob(}X\text{)}$, $\text{Prob(}Z\text{)}$, $\text{Prob(}Y\text{)}$, that are associated with the given tile are averaged and the probabilities for the logical state of each tile is provided. Then, the 4 probabilities concerning the logical state of each d=3 tile are used as the inputs of the neural network, which will provide at the output the probabilities of the logical state for the whole lattice. Based on the output of the neural network, extra corrections are going to be applied in the form of the appropriate logical operator to cancel any potential logical error created by the simple decoder. The information contained in the 32 bits of the d=3 tiles is now compressed to 16 bits that constitute the inputs of the neural network and represent the probabilities of contribution to the logical state out of every d=3 tile.

\section{Results}
\label{sec:Results}

In order to check whether the distributed decoding algorithm can reach similar decoding performance as the other popular decoding algorithms, we tested it against an unoptimized version of the Blossom algorithm \cite{edmonds, kolmogorov} and our previous implementation of neural network based decoder \cite{savvas_nn_designs} for the depolarizing error model with noiseless error syndrome measurements.

The depolarizing error model assumes errors only on the data qubits and perfect error syndrome measurements. Bit-flip (X) errors, phase-flip (Z) errors and both bit- and phase-flip (Y) errors are assumed to be generated with equal probability of $\nicefrac{p}{3}$. Such a simplistic error model is enough to prove that the distributed decoding algorithm that we propose can reach similar decoding performance to other decoding algorithms and that the scalability issues of neural network based decoder are addressed.

The critical aspect of our decoder is the choice of the size of the overlapping tiles. Since, there is only one level of concatenation, contrary to RG decoding, the size of the overlapping tiles plays a significant role in the algorithm. Having a large tile size might provide better decoding, for example decoding a d=9 surface code with d=7 tiles might be more beneficial than decoding with d=3 tiles, since there will be less shared parity-checks and long error chains will be included in a single tile. However, the bottleneck that will make such a case decode poorly in our design, is the inability of the decoder to handle properly the error syndromes unknown to the training dataset. Since it becomes exponentially harder to gather all the possible error syndromes as the code distance increases, the training dataset will be an incomplete set of all potential cases. In the case of an unknown to the training error syndrome, the neural network will not have any meaningful data to make a prediction making the behavior of the neural network inconsistent. Such a case occurs because there is an intermediate step between the cutting of the error syndrome into parts and the averaging of the probabilities of each part.

Based on that, we opted to always divide the lattice into d=3 overlapping tiles, since the d=3 case only consists of 256 different error syndromes. This is an easily obtained complete training dataset, to which any part of error syndrome of any large distance can deconstruct to. All possible error syndromes of the large lattice (d>3) are represented through the d=3 overlapping tiles, without having to explicitly sample all possible error syndromes for the large lattice.

The only downside of using d=3 tiles is that there exist some error syndromes that are highly ambiguous to what logical state they lead. Fortunately, these ambiguous error syndromes are not extremely frequent making the errors arising from this shortcoming rare. 

Another benefit of the distributed decoding approach is that the number of inputs required by the neural network is decreased compared to decoding the whole lattice approach. The reduction of inputs of the neural network for the code distances tested are shown in Table \ref{table:redInpNN}.

\begin{table}[ht]
\caption{Reduction in required inputs of the neural network}\label{table:redInpNN}
\centering
\begin{tabular}{ | c | c | c |} \hline
\text{Code distance} & \text{Old inputs} & \text{New inputs} \\ \hline
\text{d=5} & 24 & 16 \\ \hline
\text{d=7} & 48 & 36 \\ \hline
\text{d=9} & 80 & 64 \\ \hline
\end{tabular}
\end{table}

\begin{figure}[hbt]
\centering
\includegraphics[width=\columnwidth]{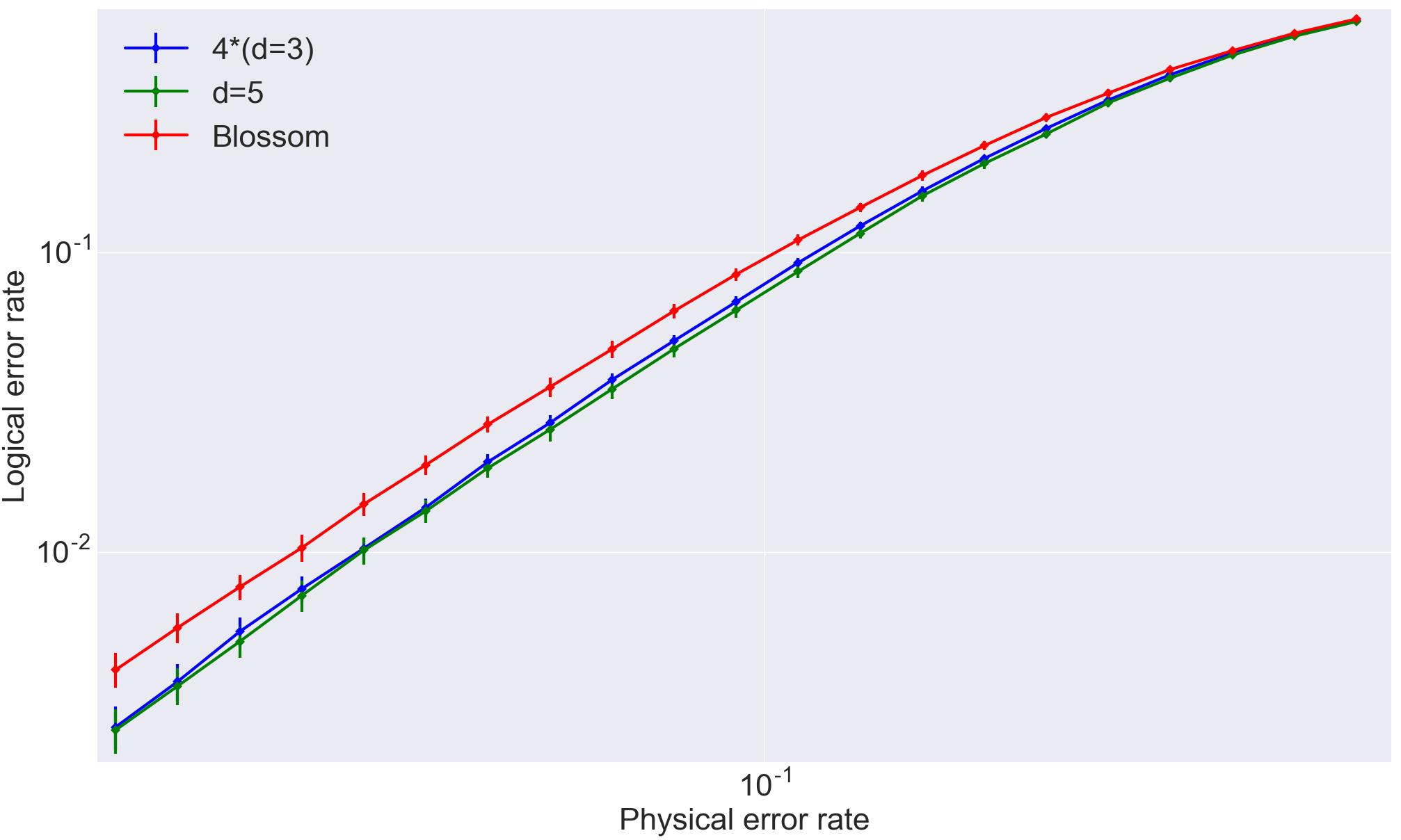}
\caption{Comparison of decoding performance between the distributed decoder with \textbf{\textit{four}} overlapping tiles of d=3 rotated surface codes inside a d=5 rotated surface code (blue), the unoptimized version of the Blossom algorithm (red) and the neural network based decoder (green)}
\label{fig:d_5_RG_4d3}
\end{figure}

\begin{figure}[hbt]
\centering
\includegraphics[width=\columnwidth]{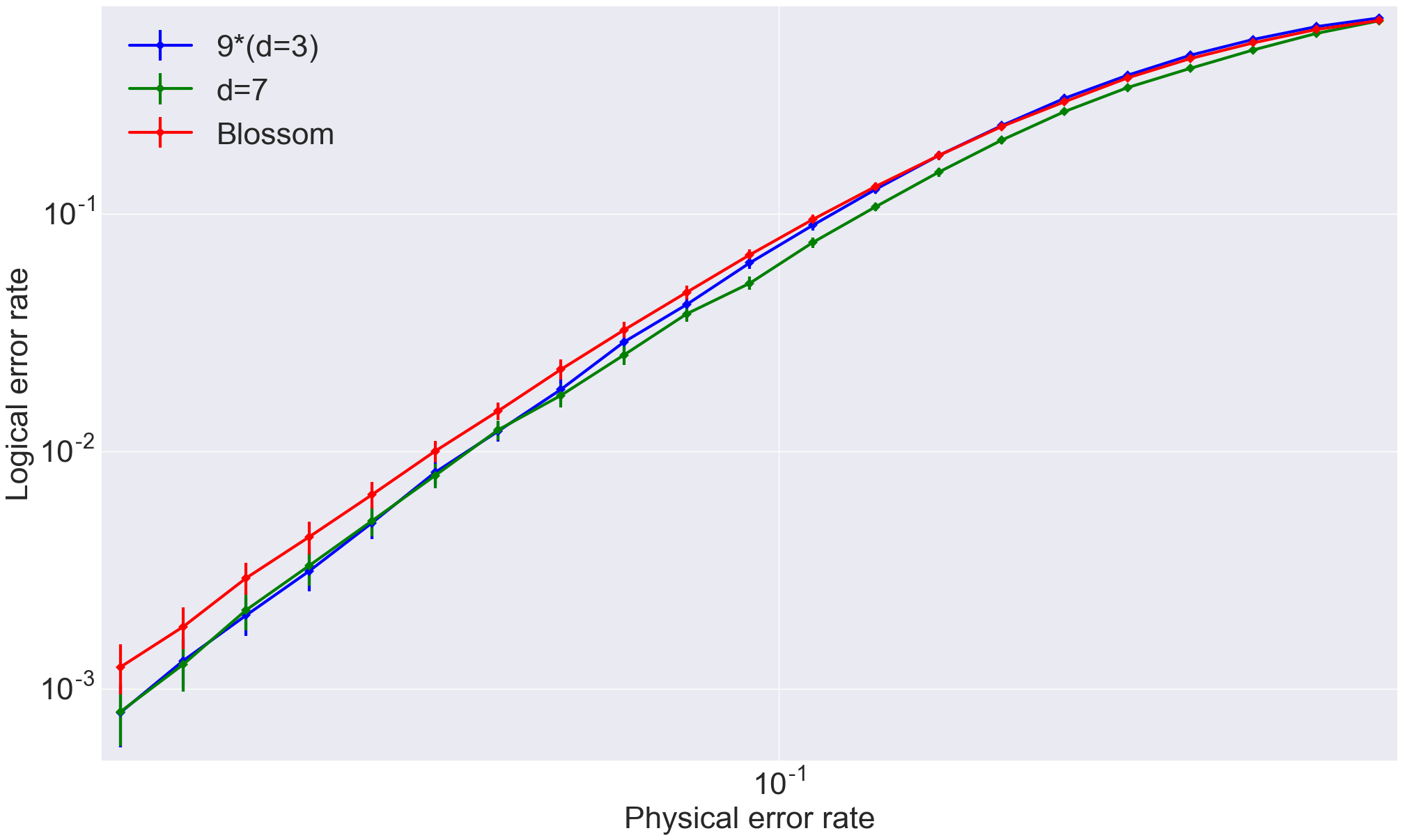}
\caption{Comparison of decoding performance between the distributed decoder with \textbf{\textit{nine}} overlapping tiles of d=3 rotated surface codes inside a d=7 rotated surface code (blue), the unoptimized version of the Blossom algorithm (red) and the neural network based decoder (green)}
\label{fig:d_7_RG_9d3}
\end{figure}

\begin{figure}[hbt]
\centering
\includegraphics[width=\columnwidth]{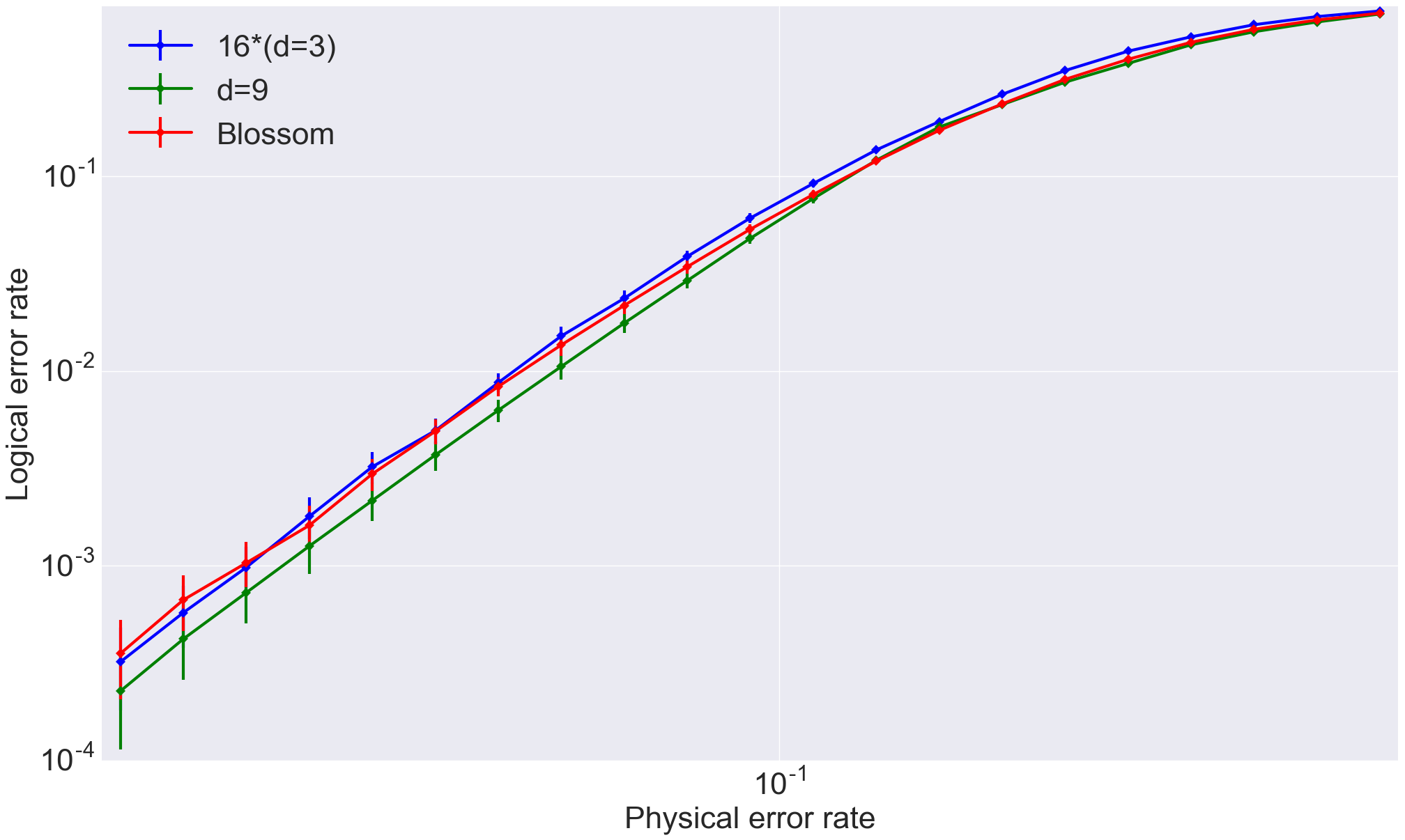}
\caption{Comparison of decoding performance between the distributed decoder with \textbf{\textit{sixteen}} overlapping tiles of d=3 rotated surface codes inside a d=9 rotated surface code (blue), the unoptimized version of the Blossom algorithm (red) and the neural network based decoder (green)}
\label{fig:d_9_RG_16d3}
\end{figure}

The comparison of the decoding performance between the distributed decoding, the neural network based decoder from \cite{savvas_nn_designs} and unoptimized version of the Blossom algorithm for a distance 5, 7 and 9 rotated surface code are presented in Figure \ref{fig:d_5_RG_4d3}, \ref{fig:d_7_RG_9d3} and \ref{fig:d_9_RG_16d3}, respectively. Each point in these graphs has a confidence interval of 99.9\%.

As can be seen from Figures \ref{fig:d_5_RG_4d3}, \ref{fig:d_7_RG_9d3} and \ref{fig:d_9_RG_16d3}, the distributed decoder can reach similar decoding performance to the compared decoders for d=5, 7 and 9, respectively. In order to have a fair comparison between the two neural network based decoders, we used the same dataset to train both decoders, therefore the decoding performance should be comparable. These comparisons were used as a proof-of-concept to verify that a distributed decoding approach is feasible and what limitations are observed.

\subsection{Optimizing for the size of training dataset}
The scalability problem that all neural network based decoders face is based on the exponential increase of the training samples required to efficiently decode. As an extension to our work on neural network based decoders, we propose an alteration to our decoding algorithm in order to increase the important training samples included in the training dataset, without increasing the size of the dataset.

As mentioned, our decoding strategy is based on a two module (simple decoder and neural network) approach, where the neural network exists to increase the decoding performance of the simple decoder. However, the simple decoder can be designed in different ways, which will lead to different decoding performance for different designs. Therefore, an investigation of the performance of the simple decoder is crucial before the training of the neural network.

\begin{figure}[htb]
\centering
\scalebox{0.62}{
\includegraphics[width=\columnwidth]{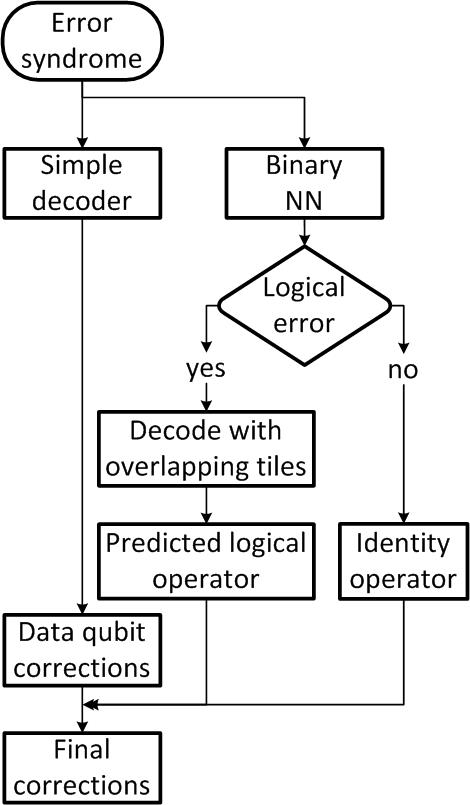}}
\caption{Description of the design flow of the optimized version of the distributed decoder}
\label{fig:optimization_flowchart}
\end{figure}

We observed that for all code distances investigated for the depolarizing error model, the simple decoder provided corrections that would lead to an error free logical state (I) \textasciitilde 42\% of the time. In those cases, the neural network would be unnecessary, since it would output the identity operator. Therefore, if we removed the error syndromes that the simple decoder corrects properly from the training dataset, then the dataset could be increased even further, with more relevant error syndromes. The only caveat is that another module, named binary neural network in Figure \ref{fig:optimization_flowchart}, should be included to the decoder which will predict whether the obtained error syndrome will be properly corrected by the simple decoder or not. The binary logic neural network might be implemented in a simpler way, which will make the binary classification task faster, instead of using a recurrent neural network as was chosen for this design.

A flowchart of the optimized algorithm with the inclusion of the extra neural network is presented in Figure \ref{fig:optimization_flowchart}. We divide the operation of the neural network from the original design of distributed decoding, to two neural networks, namely a binary neural network and a neural network for distributed decoding.

The binary neural network will predict whether the obtained error syndrome will lead to a logical error or not. The input of the binary neural network is the obtained error syndrome for the whole lattice and the output will be a binary value indicating whether extra corrections need to be applied or not. These extra corrections will arise from the neural network for distributed decoding. This neural network will work similarly to the one in the original unoptimized strategy described in section ~\ref{sec:Over_tiles}, but the training samples will be restricted to the error syndromes that lead to a logical error. The inputs and outputs of this neural network are previously explained. Note that, we need to include all 4 logical states for this neural network, because there is still a probability of an unknown to training input to produce an error free logical state.

\begin{figure}[htb]
\centering
\includegraphics[width=\columnwidth]{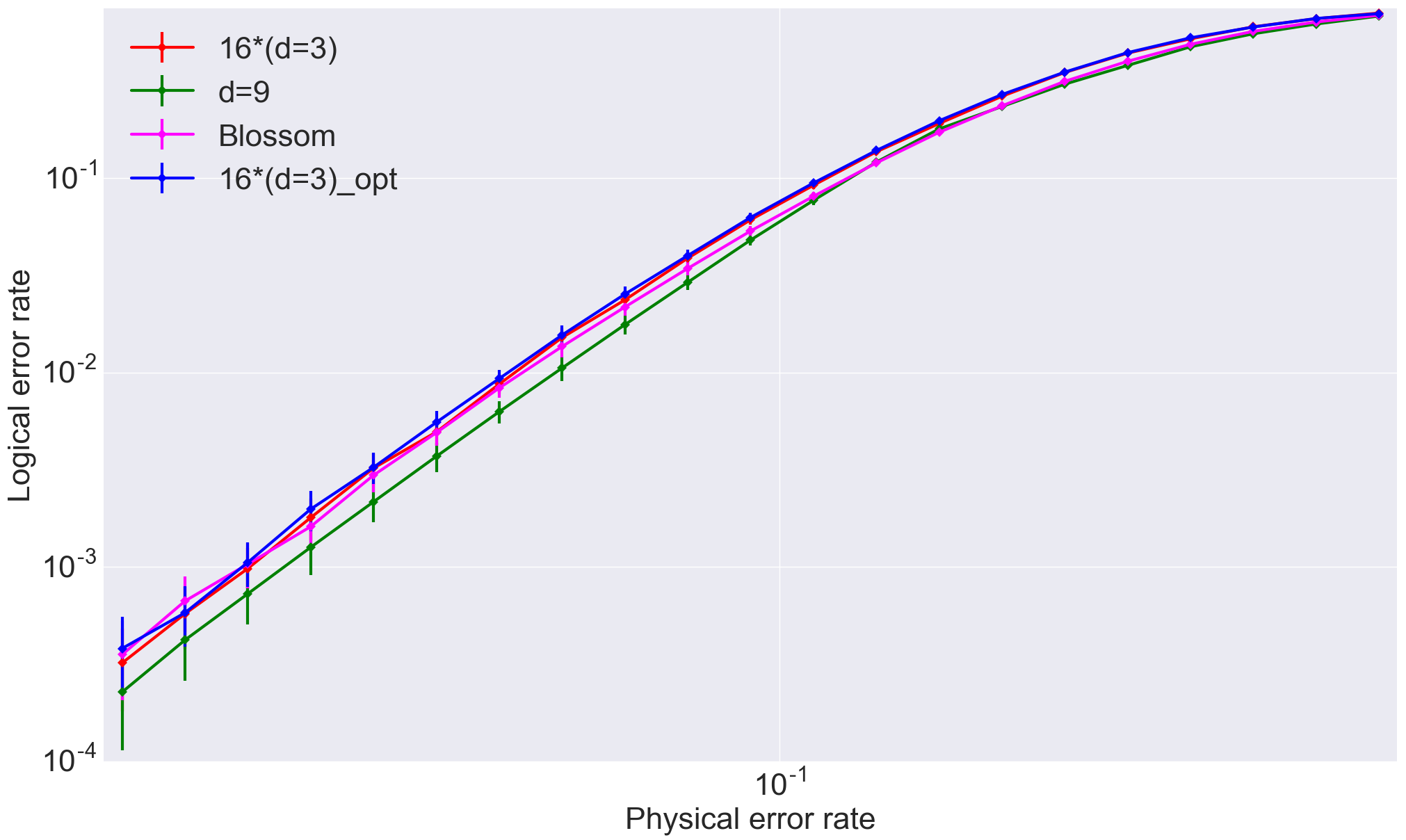}
\caption{Comparison between the optimized version of the distributed decoding (blue) to the unoptimized version (red), the unoptimized version of the Blossom algorithm (pink) and the neural network based decoder (green)}
\label{fig:d_9_all}
\end{figure}

The comparison of the decoding performance of this optimized version of the algorithm with the unoptimized one and the benchmarks that were used in this work for the largest code tested (d=9) is presented in Figure \ref{fig:d_9_all}.

As expected, the optimized version with the two neural networks cannot achieve better decoding performance than the unoptimized version, since we kept the same training dataset for both designs in order to have a fair comparison. The binary neural network has the same dataset as the unoptimized version, but the neural network for distributed decoding only includes the \textasciitilde 58\% of error syndromes that lead to a logical error. 

An important clarification is that the optimization is mentioned in the context of the potential increase of the training dataset and not in terms of better decoding performance. However, the fact that we reached the same level of decoding performance with both designs, suggests that we can make these optimizations without any loss of decoding performance.

\section{Conclusions}
\label{sec:Conclusions}

We presented a decoding algorithm that performs decoding in a distributed manner that can achieve similar decoding performance to existing decoders, like the Blossom decoder and the neural network based decoder for d=5,7 and 9. Furthermore, due to the distributed way of decoding and the deduction in the neural network inputs, larger codes can be potentially decoded. The problem of the exponential increase of the training dataset is mitigated through the distributed decoding strategy, where any error syndrome can be decomposed to smaller d=3 tiles. However, large quantum systems will still require large amounts of training samples. Moreover, in terms of execution time, we assume that a highly parallel implementation for both the simple decoder and the neural network, can potentially achieve a high speed implementation of the algorithm. Finally, we provide an alternative version of the distributed decoding strategy that can reach the same level of decoding performance as the original algorithm. The advantage of this alternative is the capability of using larger training datasets compared to other neural network based decoders, making it easier to achieve better decoding performance for higher code distances.

\def\bibsection{\section*{\refname}} 
\bibliographystyle{IEEEtran}
\bibliography{main}
\end{document}